\documentclass[conference]{IEEEtran}
\IEEEoverridecommandlockouts
\usepackage{cite}
\usepackage{amsmath,amssymb,amsfonts}
\usepackage{graphicx}
\usepackage{textcomp}
\usepackage{xcolor}

\usepackage{amsmath,amssymb,amsfonts}
\usepackage{fancybox}
\usepackage{pdfpages}
\usepackage{comment}
\usepackage{caption}
\usepackage{graphicx}
\usepackage{mathtools, cuted}
\usepackage{lipsum, color}
\usepackage{comment}
\DeclareUnicodeCharacter{00A0}{~}
\usepackage{textcomp}
\usepackage{xcolor}
\usepackage{amsmath,amssymb,amsfonts}
\usepackage{graphicx}
\usepackage{textcomp}

\def\BibTeX{{\rm B\kern-.05em{\sc i\kern-.025em b}\kern-.08em
    T\kern-.1667em\lower.7ex\hbox{E}\kern-.125emX}}
\begin{document}


\title{Invited: Human-Inspired Distributed Wearable AI\\

}

\author{\IEEEauthorblockN{Shreyas Sen and Arunashish Datta}\\
\IEEEauthorblockA{\small Elmore Family School of Electrical and Computer Engineering, Purdue University, West Lafayette, USA}\\
\IEEEauthorblockA{ \small E-mail: \{shreyas,datta30\}@purdue.edu}
}
\maketitle

\begin{abstract}

The explosive surge in Human-AI interactions, fused with a soaring fascination in wearable technology, has ignited a frenzy of innovation and the emergence of a myriad of Wearable AI devices, each wielding diverse form factors, tackling tasks from health surveillance to turbocharging productivity. This paper delves into the vision for wearable AI technology, addressing the technical bottlenecks that stand in the way of its promised advancements.

Embracing a paradigm shift, we introduce a Human-Inspired Distributed Network for Wearable AI, enabled by high-speed ultra-low-power secure connectivity via the emerging 'Body as a Wire' (Wi-R) technology. This breakthrough acts as the missing link:  the artificial nervous system, seamlessly interconnecting all wearables and implantables, ushering in a new era of interconnected intelligence, where featherweight, perpetually operating wearable AI nodes redefine the boundaries of possibility.


\end{abstract}


\maketitle

\section{Introduction}

The year 2024 is heralded as the dawn of "Wearable AI," as noted by Forbes \cite{forbes_wearable_ai}. This emergence finds its roots in the introduction of Large Language Model (LLM) based tools in late $2022$, sparking an unprecedented AI boom. With the natural language processing abilities of AI, newer and more intuitive ways of interacting with technology have proliferated. Moreover, decades of scaling semiconductor technology have culminated in a pivotal moment where significant sensing, computing, and communication power can be seamlessly integrated into miniaturized wearable devices. This convergence has paved the way for the rapid development of wearable devices empowered with Artificial Intelligence (AI). The synergy between the AI boom and the surging popularity of wearable technology has birthed a myriad of \textbf{Wearable AI} devices. From discreet pins \cite{Humane} and pocket assistants \cite{Rabbit} to elegant necklaces \cite{limitless,rewind} and immersive AR devices \cite{Meta,vision_pro, Frame}, wearable AI comes in various forms. Advancements in AI continue to blur the lines between human capabilities and machine intelligence, with wearable AI technology serving as a tangible manifestation of this convergence.

\begin{figure}[t]
\centering
\includegraphics[width=0.5\textwidth]{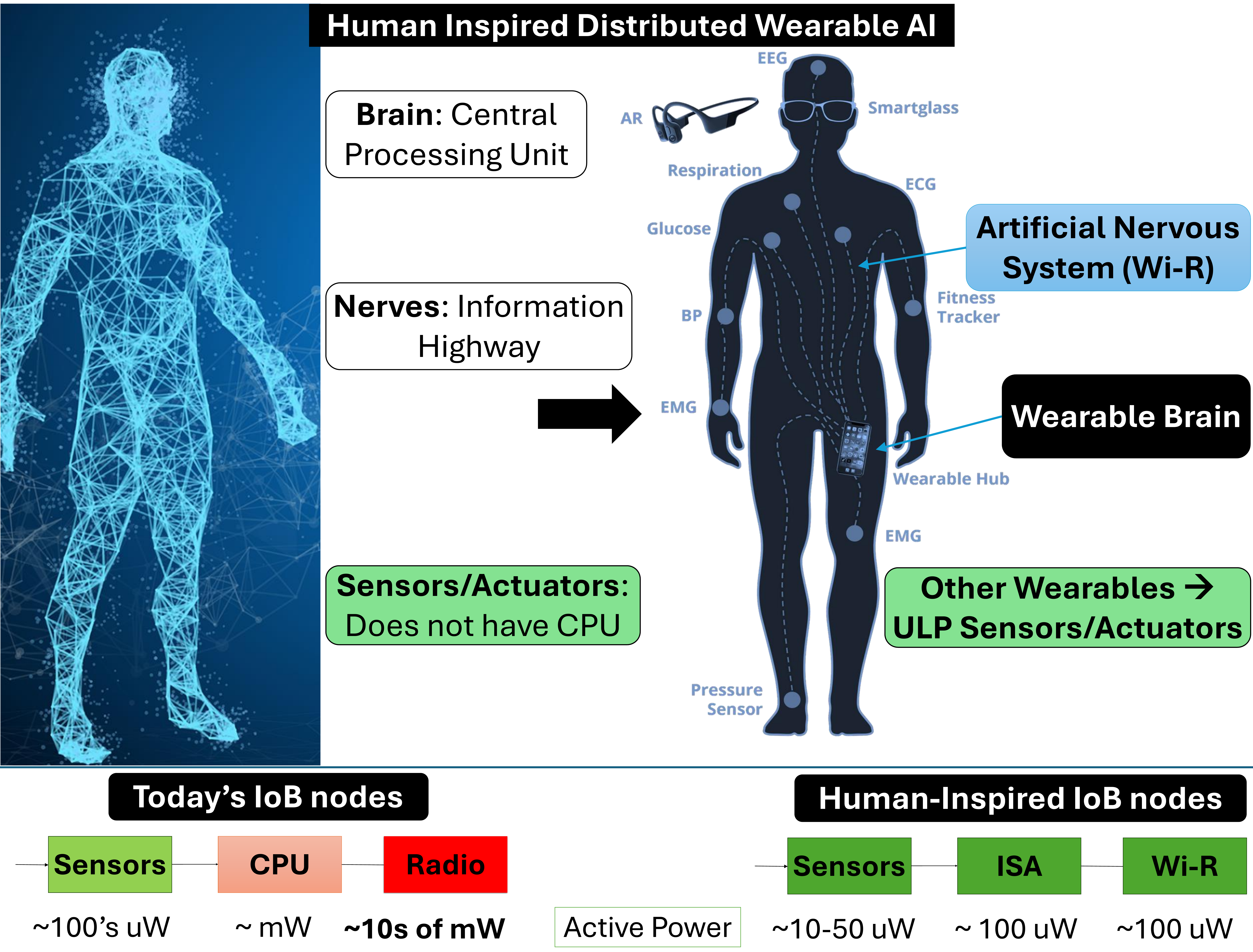}

\caption{Distributed network of wearable AI devices inspired by a centralized processing architecture found in humans.}

\label{fig:intro}
\end{figure}

The introduction of various wearable AI devices is part of a larger drive towards the exponentially increasing wearable devices over the last decade \cite{datta2023iob}. This has led to the formation of a subset of Internet of Things where the "Things" are connected by a common medium, the human body, termed as the Internet of Bodies (\textbf{IoB}) \cite{cIoB}, as shown in Fig.\ref{fig:intro}, and detailed in this IEEE Spectrum Article \cite{sen2020body}. IoB refers to network of wearables and impantables connected to an \textbf{on-body Hub}, such as a smartphone, smartwatch or a wearable brain (Fig. 1), that acts as a gateway from this network to the cloud and the internet.

To unleash the full potential of wearable AI, \textbf{IoB nodes}—sensors and actuators—must be strategically \textbf{distributed} across the body. This includes sound output near the ear, controllers near fingers or wrist, cameras on the face or chest for first-person view, and sensors like ECG near the chest, and EMG and IMU on limbs for accurate data collection. This calls for \textbf{seamleass communication} between these IoB Nodes and the On-Body Hub. 

A significant hurdle for IoB devices is the energy bottleneck, resulting in \textbf{frequent charging} and constraining wearable scalability \cite{Nasdaq}. To surmount this, there's a rising demand for charging-free, low-speed wearables enabled by energy harvesting, alongside the development of long-lasting IoB devices suited for high-speed applications like audio and video, lasting weeks to months.



It's intriguing to observe that the majority of today's wearables are equipped with a central processing unit (CPU), rendering them higher-power devices in the range of milliwatts to watts. Inspired by the architecture of human biology (Fig. 1), where distributed sensors and actuators throughout the body operate without individual dedicated CPUs, instead connected via a high-speed, low-energy nervous system to a single CPU—the brain, we ask the question: \textbf{Why can't wearable networks mimic this centralized CPU architecture found in humans?}

The answer lies in the high-energy demands of today's radio communication. It is well established \cite{chatterjee2019context,Chatterjee_Bioelectronic} that the energy consumption for radio communication per bit far exceeds that of computing per bit by several orders of magnitude. In the \textbf{absence} of a high-speed, ultra-low-power (ULP), and secure \textbf{artificial nervous system (ANS)}, today's wearables are left with no alternative but to rely on on-board computing preceding high-energy radio communication, thereby increasing platform power.

The radio communication bottleneck raises a fundamental question: \textbf{Is Radiative Communication the right technology for communicating around the conductive human body?} \cite{NSR, sen2016context}

The pursuit of an answer to this question over the past decade has spurred the development of 'Body as a Wire' or Wi-R technology \cite{maity2018bio,NSR,nath2021inter,Safety_Study,JSSC, maity2021sub,chatterjee2023biphasic}. This invention utilizes tiny, safe time-varying low to medium frequency electric fields (known as Electro-Quasistatic or EQS fields) for high-speed physically-secure communication ($>10 X$ faster than BLE) with ultra-low power consumption ($<100 X$ lower than BLE). These fields are contained around a personal bubble outside the human body, effectively creating a virtual wire that connects all wearables seamlessly. 

It's interesting to note that the human body, composed mainly of saltwater, possesses inherent conductivity, causing it to absorb radio waves. However, it remains transparent to magnetic fields and \textbf{facilitates the transmission of electric fields}, as evidenced by propagation of ECG signals around the human body. Alongside the popular radio and magnetic (e.g., NFMI) communication methods, EQS-communication emerges as a \textbf{third fundamental modality of communication}, supported by Maxwell's equations. This communication mode proves to be the most optimal for transmitting signals around conductive objects like the human body, effectively creating the missing link—the \textbf{artificial nervous system}.
Armed with seamless connectivity around the human body through Wi-R, we envision a transformative landscape (Fig. 1 right) where IoB Nodes evolve into simple sensors and actuators, along with ULP In-Sensor Analytics (ISA) as appropriate, operating at ultra-low power ($\sim 10s$ of $\mu W$ class). These nodes are all interconnected to the wearable On-Body Hub, akin to a Wearable Brain, which hosts edge intelligence and serves as a gateway to the internet. While the On-Body Hub requires daily charging, akin to current practices, the IoB nodes achieve perpetual or exceedingly long-lasting operation. This pivotal shift removes a key bottleneck of frequent charging of multiple wearables, potentially expanding the wearable market by tenfold. In doing so, it empowers humans with \textbf{real-time wearable AI through featherlight, perpetually operating IoB nodes}.

\section{Emergence of Wearable AI}
\label{sec:wearableAI}

\begin{figure}[t]
\centering
\includegraphics[width=0.40\textwidth]{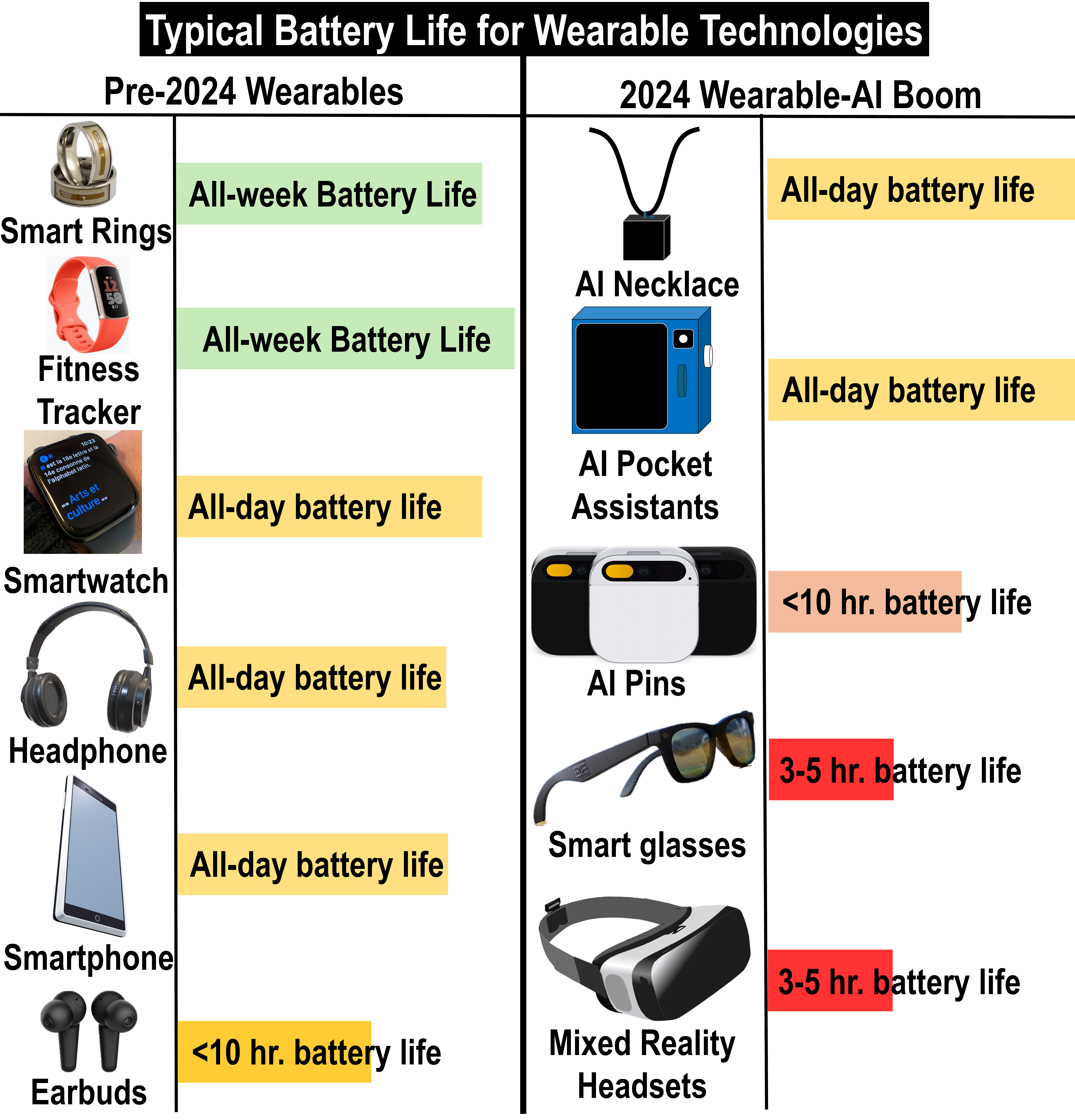}

\caption{Battery life of currently available wearable devices is illustrated. AI augmentation of pre-2024 wearables is also envisioned in the near future.}

\label{fig:wearable}
\end{figure}

Continuous semiconductor technology scaling has driven continued miniaturization of unit computing, enabling smaller and smarter wearable devices like fitness trackers, smartwatches, and smartphones. Current wearables mostly operate independently, each with its own CPU, leading to low to moderate battery life, i.e. hours to a week, depending on the size and functionality of the device (Fig. \ref{fig:wearable})). In 2024, AI capabilities have advanced to the point of real-time interactions with humans, leading to the integration of AI functionalities into a wide array of wearables, a trend expected to continue growing.

The capabilities of such wearable AI devices are enhanced when multimodal signals from distributed locations on the human body can be collected and distributed optimal actuation is performed. To fully exploit the capabilities of wearable AI, it's crucial for users to be able to seamlessly integrate these devices into their daily lives, necessitating lightweight, imperceptible designs that allow for extended comfortable use. Achieving this requires a shift in wearable architecture, with the introduction of a new layer of leaf-IoB nodes connected to an On-body Hub Edge Node, offloading heavy processing to the hub node.

We explore the current state and the future of distributed and connected wearable AI devices. While the specific form factors of these devices may evolve significantly in the years to come, the fundamental principles of distributed interconnected intelligence discussed here are expected to remain unchanged.

\subsection{Fitness trackers and health sensors}
Fitness trackers and health sensors \cite{Vivalink} are widely available wearable devices, available currently in the form factor of watches and rings, having an all-week battery life. With the advancement in ULP wearable design and energy harvesting methods, it is envisioned that miniaturized health tracking devices will become perpetually operable and can be of the form factor of wearable patches that can be worn unobtrusively around the body. 

\subsection{Voice-based devices}
With the voice interaction capabilities enabled by AI, pocket assistants of varying form factor and functionalities have emerged. The analysis of voice-based inputs and providing context dependent meaningful responses has led to the development of wearable devices like Humane AI pin, Rabbit R1, rewind and limitless AI pendants \cite{Humane, Rabbit, rewind, limitless}. Commercially available voice-operated devices currently have all-day battery life. In these devices, analyzing voice-based commands is comparatively a low power task and devices solely based on taking voice-based inputs can be envisioned to become perpetually operable in the next decade. However, voice-based responses to the commands requires higher power due to the power requirement for driving the speaker. With innovations at the circuit-level, such devices with all-day battery life are expected to be moving towards having all-week battery life.

\subsection{Image and video-based devices}
Generative AI technology has the capability to analyze and develop images and videos based on natural language inputs, as demonstrated by Dall-E \cite{DallE} and Sora AI \cite{Sora}. The use of cameras providing first-person view with devices worn on the face, as in the case of smart glasses like Ray-Ban Meta Glasses \cite{Rayban}, Frame by Brilliant Labs \cite{Frame}, mixed-reality headsets like Meta Quest \cite{Meta} and Apple Vision Pro \cite{vision_pro} or chest worn devices like Humane AI pin \cite{Humane} have paved the way for visual commands to wearable devices. These devices require low-latency computation and communication power to use video-based commands and frame appropriate responses to complete tasks real-time. Such devices interacting seamlessly with ULP nodes while having an increased battery life from all-day to all-week allows a new mode of interaction with wearables, potentially replacing existing interfaces like touchscreens.

\subsection{Neural Signals}
AI models for biopotential signals around the body (ECG, EMG) and brain signals (EEG, ECoG) have not yet gained market popularity. Providing input to Wearable AI devices using neural signals is the ultimate goal for such devices allowing the promised seamless integration of wearables into our lives. AI models recognizing neural signals will enable interaction on wearable devices without without requiring controllers, gestures or voice commands.


\section{Human-Inspired Network for IoB}
\label{sec:human-insp-BAN-IoB}
Most of today's wearable technology, equipped with standalone processing units, have power consumption in the range of $10s$ of mWs to a few watts making them too power hungry to operate perpetually. This calls for a revised architecture of distributed wearable nodes sharing resources to reduce individual power consumption. 

\subsection{Human Inspired IoB}
The human body has sensory inputs coming in from multiple points on the body to a single computing hub, the brain. The brain acts as the sole computation center, which is responsible for processing the data and communicating the response to different organs. Such a system allows ULP nodes to share resources from a central processing unit, eliminating the need for individual computational units. Thus, developing a ULP, high-speed communication method for seamless access to distributed computing is vital.

\subsection{Is RF the right technology for BAN?}
Radio frequency (RF) based radiative communication technologies have been the gold standard for wireless communication, aiding the onset of Internet of Things with smart connected devices all around us. However, RF-based communication essentially radiates the signal in a large room scale bubble around us, resulting in high power consumption of $1-10mW$. This further makes RF-based communication highly inefficient for IoB devices, as the data is radiated $5-10$ meters away from the device whereas channel lengths for IoB are typically between $1-2$ meters. This necessitates moving away from radiative communication technology for IoB nodes in order to develop ULP communication methodologies. \par
Reducing the communication power allows much of the computing to get offloaded to a remote hub. Thus, newer ultra-low-power communication techniques have been studied for wearable devices around the body which have higher energy efficiency ($\leq 100pJ/bit$), low power consumption ($\leq 100s$ of $\mu W$), and high data rates ($\geq 1Mbps$). To that end, using the body's conductive properties, treating the body as a wire \cite{sen2020body} has been explored and has been termed as Human Body Communication (HBC). \par

\section{Emergence of Wi-R as ANS}

\subsection{Electrophysiology meets Radio Comm.}
Body generated electrical signals have been known to travel through the human body at frequencies of $\leq 10 kHz$ which is firmly within the quasistatic regime \cite{NSR}. Electrophysiological recordings of such signals generated by cardiac muscles like Electrocardiogram (ECG) can now be performed using wrist worn smartwatches illustrating that these body generated signals are being communicated using the human body. We extend this concept by coupling external electro-quasistatic (EQS) digital signals to the human body at higher frequencies ($\leq 30 MHz$) to transmit data using the human body without interfering with electrophysiological signals. Even at frequencies of $10s$ of $MHz$, we observe that the properties of electrophysiological signals are consistent as these signals remain contained the human body with little to no radiation. \par
For efficient wireless communication around the human body, going to lower EQS frequencies promise higher efficiency due to higher signal absorption at radio frequencies. At EQS frequencies, a high impedance termination voltage-mode communication provides a communication channel which allows data transfer across the whole body at ultra-low communication powers. This enables externally generated digital signal communication through the body using a principle similar to that of electrophysiological signals.

\subsection{EQS Human Body Communication}
The use of quasistatic fields in communicating data has been around for over a century, with early telegraphs using Single-Wire-Earth-Return (SWER) for communicating over long distances \cite{united1966development}. In the context of Human Body Communication, operating in the quasistatic regime ($\leq 30 MHz$) provides an energy efficient and lower power alternative to RF-based communication technology \cite{zimmerman1996personal, maity2018bio}. \par
EQS-HBC has been demonstrated to be a physically secure \cite{NSR}, low power ($\approx 415 nW$ for $10kbps$) \cite{maity2021sub} and energy efficient (Sub-$10pJ/bit$) \cite{JSSC} solution for communication between IoB devices. Wi-R, a commercial implementation of EQS-HBC has been demonstrated to show high data rate ($4 Mbps$) communication with an energy efficiency of $\approx 100 pJ/bit$ \cite{datta2023can, senWiR}. Future research in HBC is focused on increasing communication data rate while ensuring high energy efficiency and exploring body-assisted communication for implantable devices in EQS regime and beyond using Magneto-Quasistatic Human Body Communication leveraging the human body’s transparency to magnetic fields. \par

\begin{figure}[t]
\centering
\includegraphics[width=0.5\textwidth]{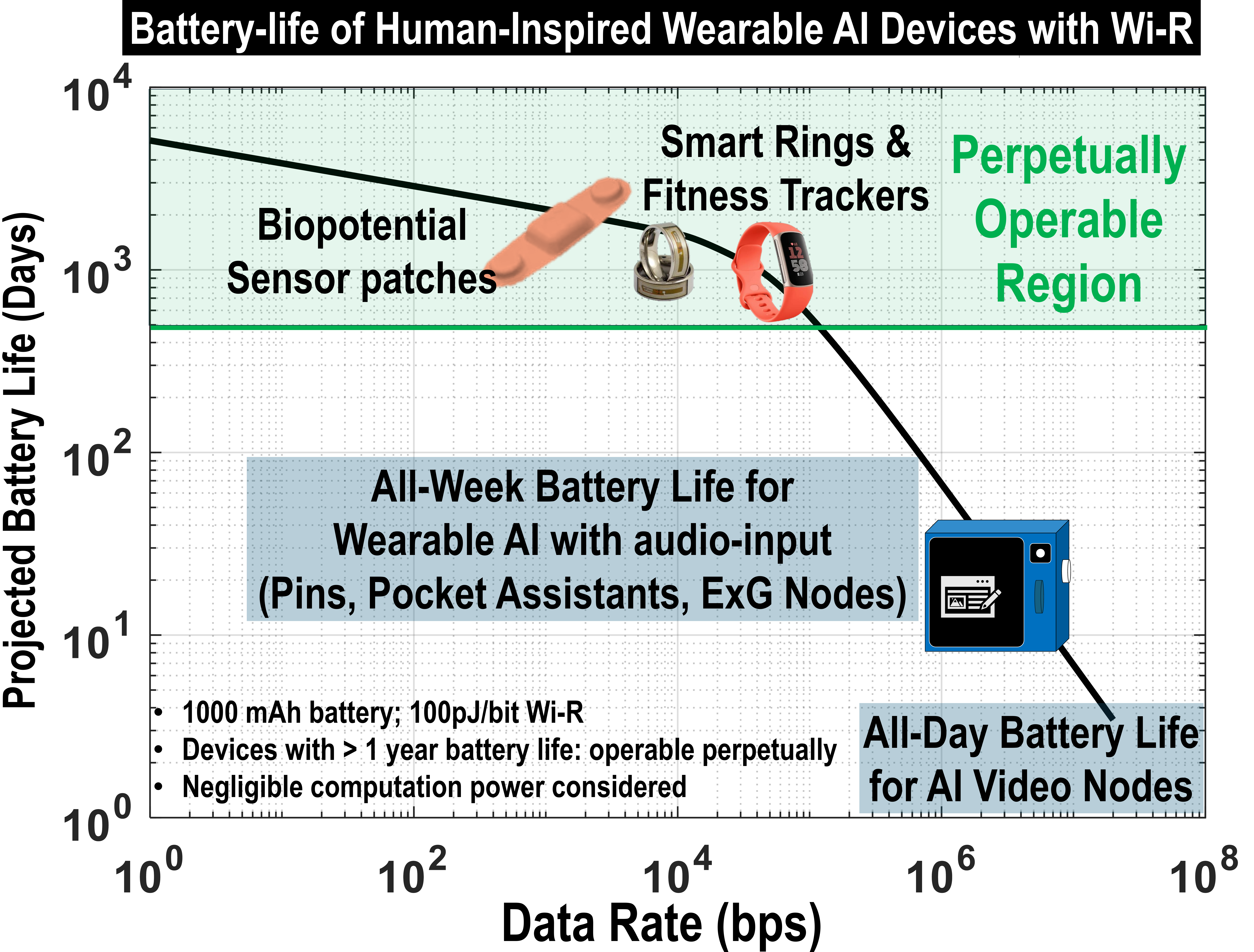}

\caption{Projected battery life of wearables with respect to data rate using Wi-R \cite{datta2023can}. Power consumption is calculated using a survey of literature and commercial devices \cite{datta2023can}.}

\label{fig:Battery_life_WiR}
\end{figure}

\section{Distributed IoB Wi-R Network}
With Wi-R, an ultra-low-power communication methodology can be implemented for Body Area Networks (BAN), allowing a distributed Wearable AI computing platform. Leaf nodes, which are the ultra-low-power wearables in the IoB architecture, are connected to edge (hub) which consist of larger devices like mixed reality headsets and smartphones having higher available computational power. This allows the creation of perpetually operating wearables which use the computational resources of the hub to perform power hungry tasks using ultra-low-power communication enabled by Wi-R. The ULP nodes in some cases may use low power in-sensor analytics (ISA) or data compression (example MJPEG compression for video) to reduce the data volume to be communicated. The hubs are connected to fog and cloud servers for further data analytics.\par
Fig. \ref{fig:Battery_life_WiR} illustrates battery life for wearable nodes communicating using EQS-HBC for a battery capacity of $1000 mAh$ which can be achieved with a high capacity coin cell battery \cite{Github_graph}. Although ISA may be used in the ULP nodes, this approximation considers the compute power to be negligible as a first order approximation considering the total power to be the sum of sensing and communication power consumption. We calculate the communication power consumption for Wi-R with an energy efficiency of $100pJ/bit$ \cite{senWiR}. The sensing power is characterized as a function of data rate with a survey of past literature and commercially available analog front-ends \cite{datta2023can}. We further consider devices with more than a year of battery life as perpetually operable. With current energy harvesting modalities, $10-200 \mu W$ power harvesting is possible in indoor conditions. Using Wi-R to communicate between leaf and edge nodes, it is projected that wearable devices like biopotential sensors, smart rings and fitness trackers can be made perpetually operable. Further, audio and video nodes with low computation power can be made all-week and all-day operable respectively. 

\section{Conclusion}

The convergence of human-AI interactions and wearable technology has thrust Wearable AI devices into the spotlight of innovation in 2024, revolutionizing fields from healthcare to productivity enhancement. This paper explores the future trajectory of Wearable AI, tackling obstacles to its widespread adoption. We advocate for a human-inspired distributed network model, linking wearable AI leaf-nodes to an on-body hub via Wi-R (Body as a Wire) technology. This approach facilitates lightweight, perpetually operable wearable AI devices by shifting intensive computing tasks to the edge hub, eliminating the inconvenience of frequent charging and enabling seamless integration into daily human life.

\bibliographystyle{IEEEtran}

\bibliography{references}

\end{document}